\documentclass{v17cosmo}

\def\Journal#1#2#3#4{{#1} {\bf #2}, #3 (#4)}

\def\PRD{{\em Phys. Rev.} D}
\def\aap{\em Astron.~Astrophys.}

\def\jcap{\em J.~Cosmol.~Astropart.~Phys.}
\def\mnras{\em Mon.~Not.~R.~Astron.~Soc.}
\def\apj{\em Astrophys.~J.}

\def\be{\begin{equation}}
\def\ee{\end{equation}}
\def\bea{\begin{eqnarray}}
\def\eea{\end{eqnarray}}

\newcommand{\Photo}{\includegraphics[height=35mm]{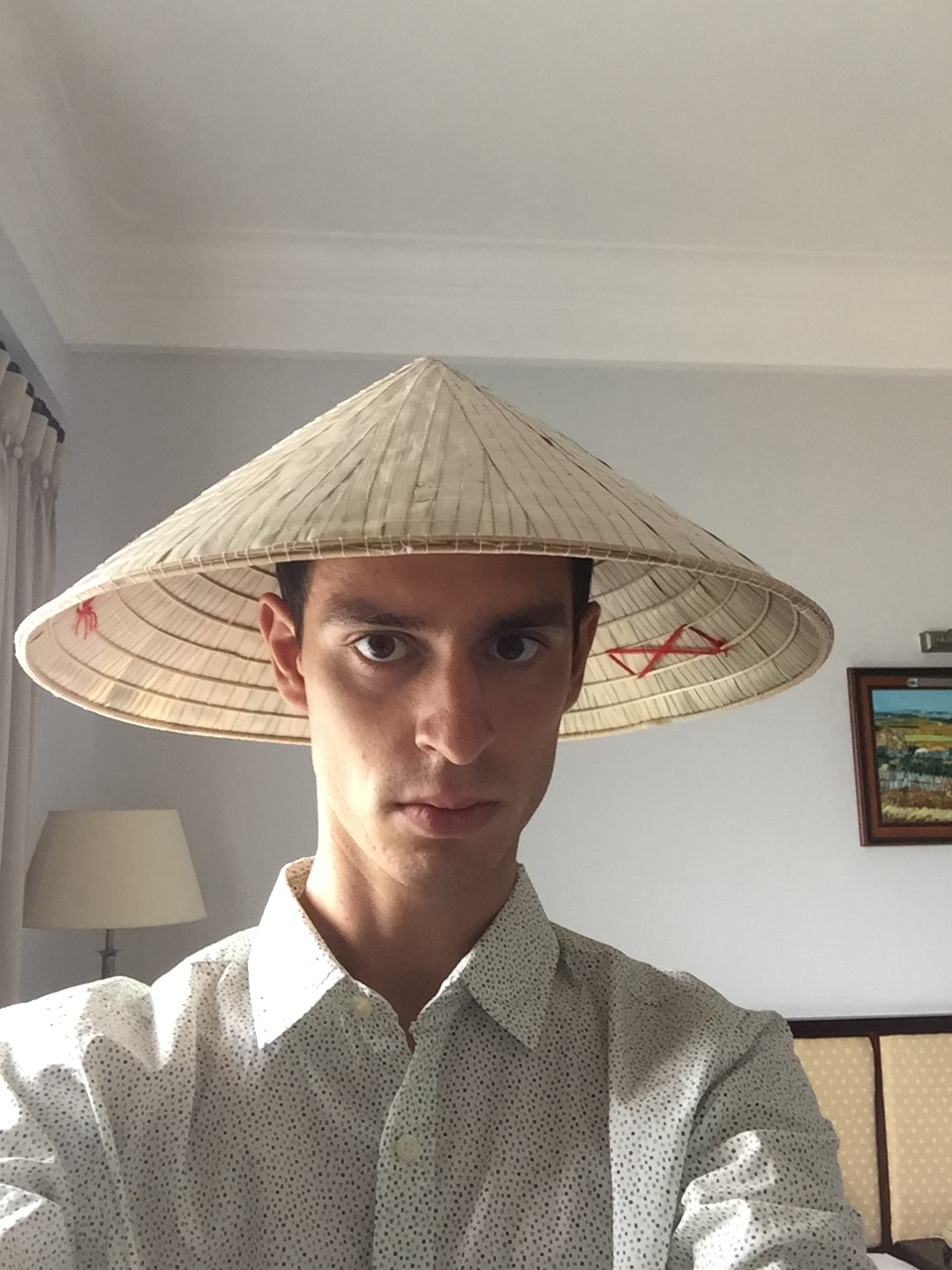}}

\begin{document}
\vspace*{4cm}
\title{BISPECTRUM MODELLING IN LARGE SCALE STRUCTURE - A THREE SHAPE MODEL}

\author{ A. LAZANU }

\address{INFN, Sezione di Padova, Via Marzolo 8,\\
I-35131 Padova, Italy}

\maketitle\abstracts{
We study the matter bispectrum of large structure by comparing theoretical models (perturbation theories and halo models) to numerical simulations using shape and amplitude correlators. We show that among the perturbation theories at one loop the effective field theory of large scale structure extends the furthest into the non-linear regime. We analyse the one and two-loop bispectra in the renormalised perturbation theory and we show that there is a significant extension in the range where results are accurate when going to two loops. In the case of the halo model, we show that there are deficiencies in the modelling of the two-halo term at redshifts $z>0$ that worsen in the past. Based on this observation and on the shapes identified in the halo model, we build a simple `three-shape model' that provides a good fit to $N$-body simulations on all scales, at both low and high redshifts. We show that this model can be easily extended to local and equilateral primordial non-Gaussianity using the same shapes.
}

\section{Introduction}\label{sec:intro}
Cosmic Microwave Background (CMB) probes, and in particular \textit{Planck}, have provided a wealth of information in recent years, confirming that the six-parameter $\Lambda$CDM model and the inflationary paradigm provide an adequate description of the observed Universe \cite{2015arXiv150201582P}. Many theories have been developed to explain the mechanism of inflation. Most of them predict the same two-point correlation function (power spectrum) of the temperature anisotropy, but higher-order correlations may be different. Therefore, in order to distinguish between different models, one must look at these correlators, starting with the three-point function (bispectrum).
Complementary information to the CMB can be obtained from the late-time distribution of galaxies (large scale structure -- LSS). This contains significantly more information than the CMB, due to its three-dimensional nature, but it is much more difficult to extract, because of the non-linearities, redshift space effects and the relationship between the observed galaxies and the underlying dark matter distribution (bias). 

In this paper, we study the modelling of the matter bispectrum of LSS by analysing predictions of theoretical models. We compare numerical bispectra arising from perturbation theories and halo models with $N$-body simulations, in the absence \cite{bislong} and in the presence of primordial non-Gaussianity (PNG) \cite{2017PhRvD..95h3511L} using an innovative three-dimensional approach based on shapes, and we build a new `three-shape' model that provides a very good fit to the simulations.

\section{Matter Bispectrum and Correlators}
By denoting the matter overdensity $\delta$ we define its power spectrum and bispectrum in Fourier space as
\begin{equation}
\langle	\delta (\textbf{k}_1) \delta (\textbf{k}_2) \rangle = (2\pi)^3 \delta_D (\textbf{k}_1 + \textbf{k}_2) P(k) \, , 
\end{equation}
\begin{equation}
\langle	\delta (\textbf{k}_1) \delta (\textbf{k}_2) \delta (\textbf{k}_3) \rangle = (2 \pi)^3 \delta_D (\textbf{k}_1 + \textbf{k}_2 + \textbf{k}_3) B(k_1,k_2,k_3) \, ,
\end{equation}
where $\delta_D$ is the Dirac delta function and $k_i=|\textbf{k}_i|$. We define the scalar product between two bispectra $B_i$ and $B_j$ as
\begin{equation}
\label{shapeprod}
\langle B_i, B_j \rangle \equiv \frac{V}{\pi}\int_{\mathcal{V}_B}dV_k\, \frac{k_1k_2k_3 \,B_i(k_1,k_2,k_3)\,B_j(k_1,k_2,k_3)}{ P(k_1)P(k_2)P(k_3)} \ ,
\end{equation}
where $V$ is the volume of integration and $\mathcal{V}_B$ represents the region defined by the triangle condition on wavevectors $\textbf{k}_1$, $\textbf{k}_2$, $\textbf{k}_3$. We consider two regions for $\mathcal{V}_B$: (i) $k_1, k_2, k_3 \le k_{max}$ -- the cumulative scalar product; (ii) $K-\Delta K \le k_1 +k_2 +k_3 \le K +\Delta K$ -- the sliced scalar product. Using Eq.~\ref{shapeprod}, we define shape ($\mathcal{S}$) and amplitude ($\mathcal{A}$) correlators between two bispectra
\begin{eqnarray}
&\mathcal{S}\left(B_i,B_j\right) \equiv {\langle B_i, B_j \rangle}/{\sqrt{\langle B_i, B_i \rangle \langle B_j, B_j \rangle}} \, , \\
&\mathcal{A} \left( B_i, B_j \right) \equiv  \sqrt{ {\langle B_i, B_i \rangle}/{\langle B_j, B_j \rangle}} \, .
\end{eqnarray}
These two correlators can be used to compare how much two bispectra resemble, and in particular they can be employed to compare theoretical models with simulations, to find up to which scale one can expect a theoretical model to be accurate (cumulative scalar products), or to check resemblances on scale-invariant slices (sliced scalar products). For a model to be an accurate description of data, both shape and amplitude correlators must be as close as possible to unity.

\section{Simulations}
For analysing the performance of the theoretical models, we have used $N$-body simulations \cite{Schmittfull2013} based on modal estimators \cite{Fergusson2012,Regan2012} with both Gaussian and non-Gaussian initial conditions ($f_{NL}^{\mathrm{loc}}=10$ and $f_{NL}^{\mathrm{equilat}}=100$), each of them having three realisations. Around 100 coefficients have been required to accurately reconstruct the matter bispectrum from the modal expansion. In the case of Gaussian initial conditions, we have smoothly combined three simulations \cite{bislong}, covering scales up to $k \sim 7.8 h/\textrm{Mpc}$ and for the non-Gaussian scenarios we have employed two sets of simulations up to $k \sim 2.0 h/\textrm{Mpc}$. Each of the simulations considered contains $512^3$ particles, the initial redshift is 49 and the box sizes are $1600 (h/\textrm{Mpc})^3$, $400 (h/\textrm{Mpc})^3$ and $100 (h/\textrm{Mpc})^3$ respectively.

\section{Theoretical Models}
\subsection{Perturbation Theories}
\label{subsec:pt}
Perturbation theories are describing matter clustering on mildly non-linear scales and represent corrections to linear theory.
In this work we focus mostly on one-loop results, and we also show an extension to two-loops. We have analysed the following models in perturbation theory:

\begin{itemize}
\item 
The Eulerian \textit{Standard Perturbation Theory} (SPT) is based on a linearised expansion of the evolution equations for the overdensity and velocity fields \cite{Fry1984,Bernardeau20021}, when these are much smaller than unity. In the case of the bispectrum, the lowest order non-zero term is called the `tree-level' bispectrum. At the next order (one loop), there are four non-zero terms. This theory is however inaccurate, because the loop integrals involve integration over infinite domains, where the assumption $|\delta| \ll 1$ is no longer valid. Moreover, there are cancellations between terms which does not guarantee a more accurate result at a larger number of loops. The bispectrum predicted is too high compared to numerical simulations. Various approaches to resolve this issue have been proposed, some of which are summarised in the next few paragraphs. 

\item
The \textit{Effective Field Theory} (EFT) of LSS \cite{1475-7516-2012-07-051,Carrasco2012} modifies the evolution equations used in SPT, by considering terms that account for the effect of short wavelength modes on the long wavelength ones. These induce corrections to the linearised fluid equations, and in the case of the bispectrum, all the of SPT terms are recovered, but also additional correction terms that must be added. The counterterms subtract the excessive SPT contribution, increasing the accuracy of the modelling \cite{Angulo,baldauf}. This theory requires fitting parameters to numerical simulations for the power spectrum, and the same parameters can be used for the bispectrum. 

\item The \textit{Renormalised Perturbation Theory} \cite{PhysRevD.73.063519,Bernardeau2008} recasts the evolution equations used in SPT in matrix form and solves them using a `non-linear propagator'. In this work we use its simplified form \textsc{MPTbreeze} \cite{Crocce2012}. This yields a resummed, convergent expansion that becomes more accurate at a higher number of loops. Technically, only some of the terms of SPT are required. We have computed numerically the bispectrum of this theory at both one and two loops.

\item \textit{Resummed Lagrangian Perturbation Theory}.
One can derive the equations of motion of the fluid displacements in Lagrangian coordinates, and a resummed expression for the bispectrum can be obtained \cite{Matsubara2008,Rampf}.

\end{itemize}
\subsection{Halo Models}
The halo model is a phenomenological model, based on the spherical collapse model, that can be used to extend clustering to non-linear scales. The main assumption is that all the matter present in the Universe is contained within virialised haloes. The model has three main ingredients: the halo profile, the halo mass function and the bias functions \cite{Seljak11102000,Cooray20021}. In this model, the bispectrum is expressed as a sum of three terms -- the one-, two- and three-halo terms, describing contributions of three particles in the same halo, two in one halo and another in a different halo, or all three in different haloes.  Although the halo model provides a useful description of matter clustering on all scales, it has two deficiencies: (a) on large scales, at late times, there is an excess of power because the one and two-halo terms do no decay fast enough; (b) on intermediate scales, there is a deficit of power at early times, because the assumption that all matter is contained within virialised haloes is increasingly inaccurate at higher redshifts. One solution to resolve the problems on large scales has been proposed in Refs. \cite{valageas1,valageas2}, where a perturbative method is combined with the halo model, effectively replacing the three-halo term. In our implementation, we have used EFT as the perturbation theory.

\section{Primordial non-Gaussianity}\label{sec:png}
When initial conditions are non-Gaussian, the matter bispectrum gets modified according to the shape of the primordial bispectrum, which is based on the inflationary model considered. Its amplitude is quantified through the parameter $f_{NL}$. In this work, we will only focus on the local and equilateral shapes. For perturbation theories, there are additional terms that must be added to the Gaussian expansion \cite{2010MNRAS.406.1014S}; in particular, there is an additional tree-level contribution. In the case of EFT, counterterms must be added to account for the effect of the short wavelength modes, as for the Gaussian scenario \cite{2015JCAP...11..024A}. For the halo model there are modifications to the profile, mass function and bias \cite{1475-7516-2012-08-036}. The effect of PNG is expected to be small compared to the Gaussian component. Therefore, in order to make quantitative statements about the non-Gaussian contribution, we have subtracted the Gaussian component of the bispectrum and we analysed only the non-Gaussian remainder [$\Delta B_{NG} \equiv B(f_{NL})-B(f_{NL}=0)$].

\section{Bispectrum Shapes}\label{sec:shapes}
We consider the following shape functions:
\begin{equation}
\label{shtree}
S^{\mathrm{tree}}(k_1,k_2,k_3)=2\left[F_2^{(s)}(\mathbf{k}_1,\mathbf{k}_2)P_{\mathrm{lin}}(k_1)P_{\mathrm{lin}}(k_2) + 2 \,\mathrm{perms} \right] \, ,
\end{equation}
\begin{equation}
\label{shsq}
S^{\mathrm{squeezed}}(k_1,k_2,k_3)=\frac{1}{3} \left[P_{\mathrm{lin}}(k_1)P_{\mathrm{lin}}(k_2) + 2 \,\mathrm{perms} \right] \, ,
\end{equation}
\begin{equation}
\label{shcon}
S^{\mathrm{constant}}(k_1,k_2,k_3)=1 \, (\mathrm{Mpc}/h)^6 \, ,
\end{equation}
where Eq. \ref{shtree} is the gravitational tree-level bispectrum.
\begin{figure}
\begin{center}
\includegraphics[height=5 cm]{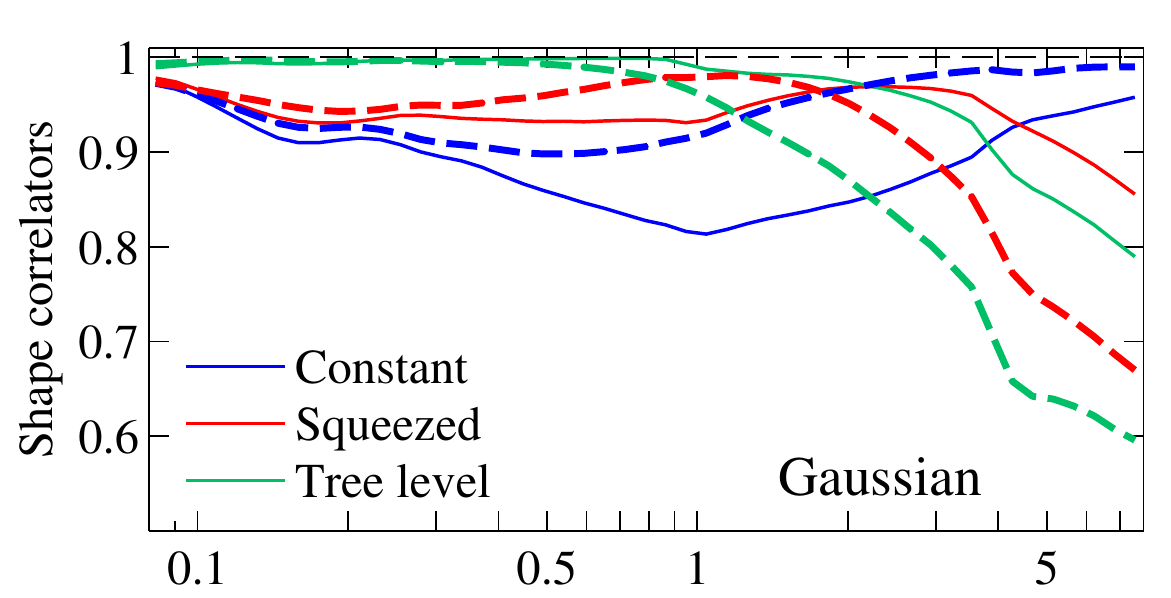}
\caption{Shapes of the Gaussian $N$-body simulations. The thick dashed lines represent $z=0$ and the narrow continuous line -- $z=0$. The shapes at $z=1$ follow a similar pattern, but are omitted for simplicity.}
\end{center}
\end{figure}
By computing the sliced shape correlators between the perturbation theories at one loop perturbation and the three shapes described above (Fig. 6 of Ref. \cite{bislong}), we show that the perturbative theories have a flattened shape, as the gravitational tree-level bispectrum. In a similar manner, the three terms of the halo model each have different shapes -- the one-halo term has a constant shape, the two-halo term has a squeezed shape, while the three-halo term has a flat shape (Fig. 7 of Ref. \cite{bislong}). The three shapes can also be observed in the simulations (Fig. 1): on large scales, in the perturbative regime, the bispectrum has a flattened shape, corresponding to the three-halo term; on intermediate scales, where the two-halo term is dominating, there is a squeezed shape; on small scales, in the deeply non-linear regime, the bispectrum has a constant shape corresponding to the one-halo term. Moreover, in Fig. 1 the evolution of the shapes with redshift is evidenced: the flattened shape becomes more extended as the redshift is increased (as structures become more linear at earlier times and linear theory is becoming an increasingly good approximation), while the squeezed and constant shapes shift towards non-linear scales.

\section{The Three Shape Model}
Based on the observations from the halo model, we propose the following decomposition for the matter bispectrum:
\begin{equation}
\label{decomp}
 B(k_1,k_2,k_3)=\sum_{i=1}^3 f_i(K)S^i(k_1,k_2,k_3) \,
\end{equation}
where $K=k_1+k_2+k_3$, the shape functions $S^i$ are given by Eqs. \ref{shtree}-\ref{shcon}, and functions $f_i$ are unknown amplitude functions. We show that this ansatz can yield an improved fitting to the data with respect to the halo model on all scales.

On small scales, the bispectrum has a constant shape. We have seen that this model accurately describes matter clustering. Therefore, we choose to fit the amplitude function to the one-halo term and use it directly in our three-shape model. Due to the particular form of the ansatz, we can use the equilateral component of the model to find a fitting function of the form,
\begin{equation}
f_{\mathrm{constant}}(K)=\frac{A}{\left(1+bK^2 \right)^2} \, ,
\end{equation}
where the two coefficients are then easily obtained for each redshift. 

On intermediate scales, the halo model is no longer accurate at redshifts $z>0$ and moreover the two-halo term does not decay fast enough at low $k$; hence, we choose to use a function inspired by the two-halo term, but we fit the coefficients to the simulations:
\begin{equation}
f_{\mathrm{squeezed}}(K)=\frac{C}{\left(1+DK^{-1} \right)^3} \,.
\end{equation}

\begin{figure}
\begin{center}
\includegraphics[height=6 cm]{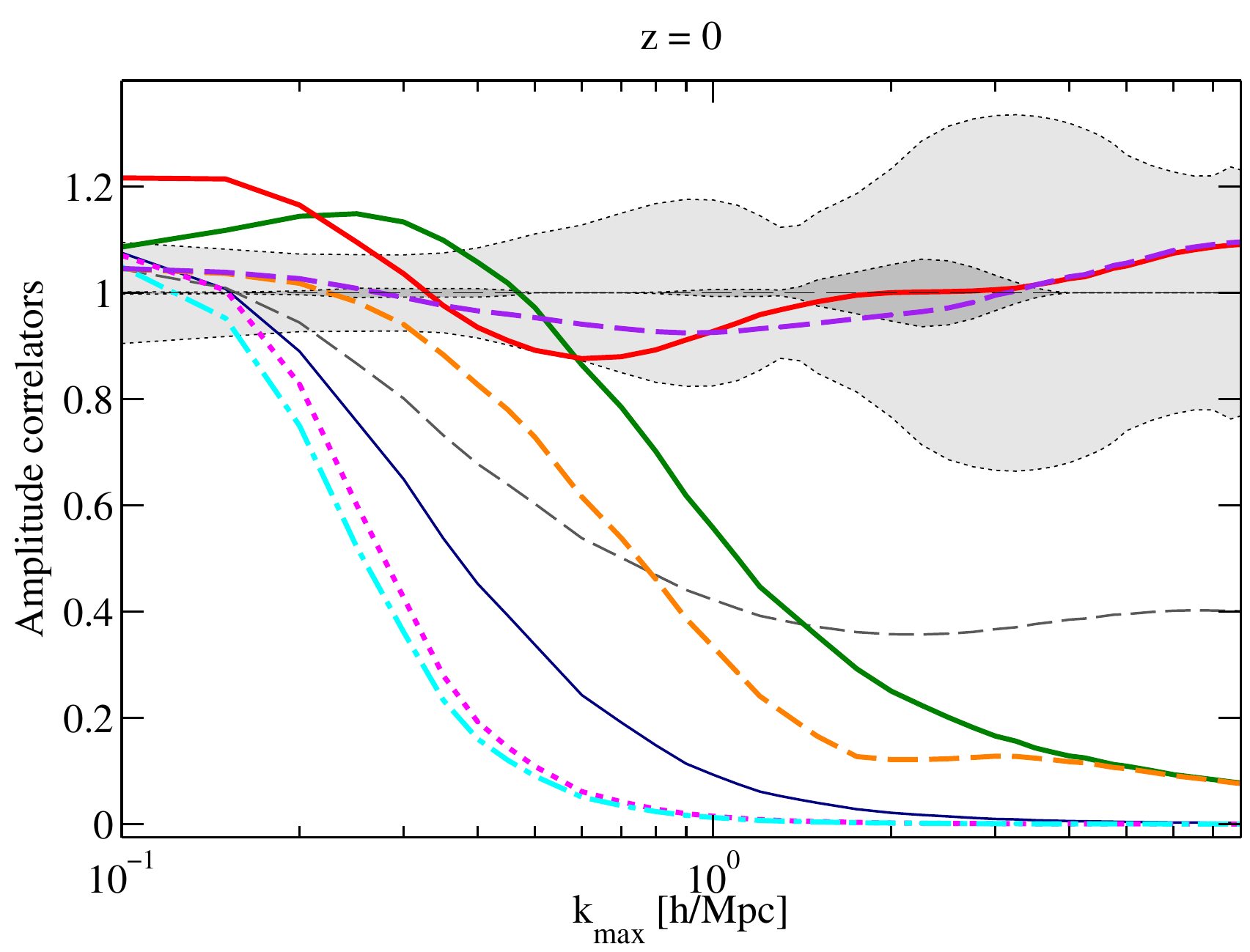} \includegraphics[height=6 cm]{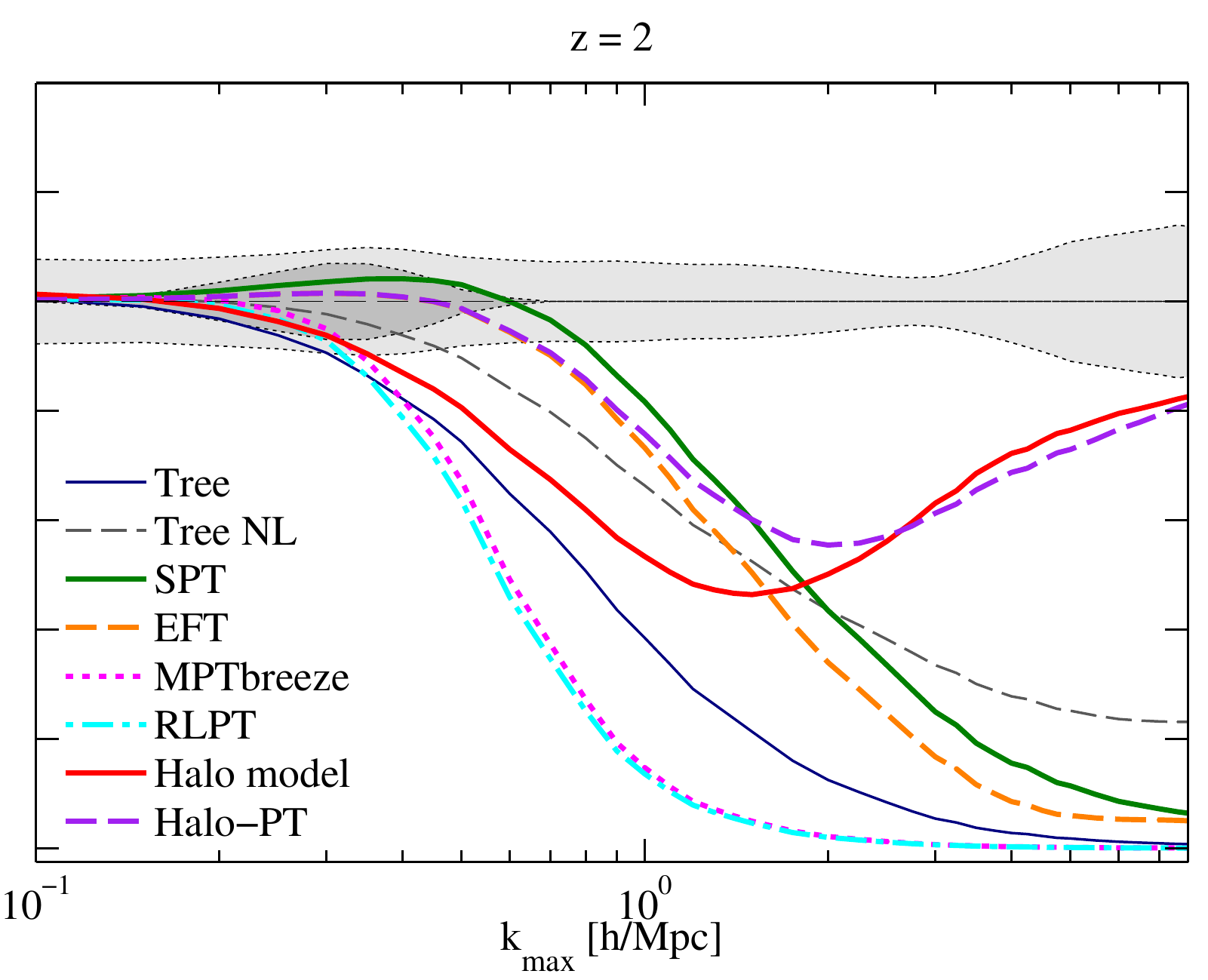} \\
\caption{Amplitude correlators between theoretical models and the three-shape model (for the Gaussian case) at $z=0$ (left) and $z=2$ (right). The legend is common to both plots.}
\end{center}
\end{figure}

On large scales, perturbation theories are providing accurate descriptions of the matter bispectrum. However, all one-loop perturbation theories are flattened \cite{bislong}; hence the tree-level bispectrum is not sufficient because it quickly decays. In order to keep the model simple, we choose to use the gravitational tree-level bispectrum, but replace the linear power spectrum in Eq. \ref{shtree} with the \textsc{HALOFIT} power spectrum \cite{Smith,2012ApJ}, thus creating a `non-linear tree-level' bispectrum. This provides excessive signal on small scales (Fig. 2) and hence we cut it off with an exponential of the form $f_{\mathrm{tree}}(K)=\exp(-K/E)$. Coefficients $C$, $D$ and $E$ are then fitted to the simulations, thus determining the three-shape model. 

In the case of the non-Gaussian bispectrum, the ansatz from Eq. \ref{decomp} remains largely unchanged, with the exception that the tree-level shape (Eq. \ref{shtree}) is replaced with its non-Gaussian counterpart \cite{2017PhRvD..95h3511L} (which depends on the shape of PNG),
\begin{eqnarray}
\Delta B_{NG}^{3-\mathrm{shape}}(k_1,k_2,k_3)=S^{\mathrm{tree}}_{NG,NL}+c_1f_{\mathrm{squeezed}}(K)S^{\mathrm{squeezed}}(k_1,k_2,k_3) \nonumber \\
+c_2f_{\mathrm{constant}}(K)S^{\mathrm{constant}}(k_1,k_2,k_3) \, .
\label{decompng}
\end{eqnarray}
The amplitude functions also remain the same, but they are multiplied by different numerical coefficients ($c_1$, $c_2$), which are fitted to simulations.

Both the three-shape model (Eq. \ref{decomp}) and its non-Gaussian correction (Eq. \ref{decompng}) provide a good fits to the $N$-body simulations, as the shape and amplitude correlators between the three shape model and the simulations are close to unity \cite{bislong,2017PhRvD..95h3511L}.

\section{Results}

We have compared the predictions of theoretical models -- perturbation theories and halo models with the three-shape model and numerical simulations using  the cumulative shape and amplitude correlators. We have represented fitting errors of the three shape model in dark grey, while the light grey areas represent uncertainties between realisations of the simulations.

\subsection{Perturbation theories (one loop) and halo models}
We have investigated the predictions of perturbation theories at one loop by looking at the shape and amplitude correlators. The shape correlators do no represent an accurate method of testing the predictions because they are always smaller or equal to unity and hence one cannot distinguish between the situations when the model is underestimating or overestimating the actual bispectrum. Amplitude correlators are however more promising. In Fig. 2 (left) we have plotted the amplitude correlator between the various theoretical models and the three shape model at redshift zero for the Gaussian scenario. The plot shows that SPT is overestimating the signal on mildly non-linear scales and that the EFT counterterm subtracts the effective contribution. The \textsc{MPTbreeze} and RLPT (both one loop) predict a similar signal that is exponentially suppressed. The `non-linear' tree-level bispectrum, built using the \textsc{HALOFIT} power spectrum has a significant leftover signal on small scales, and in order to avoid any excessive flat signal we have cut it off with the exponential function in the three-shape model. The plot shows that the EFT extends the furthest into the nonlinear regime and this trend is also valid at earlier times (Fig. 2 - right).

\begin{figure}
\begin{center}
\includegraphics[height=6 cm]{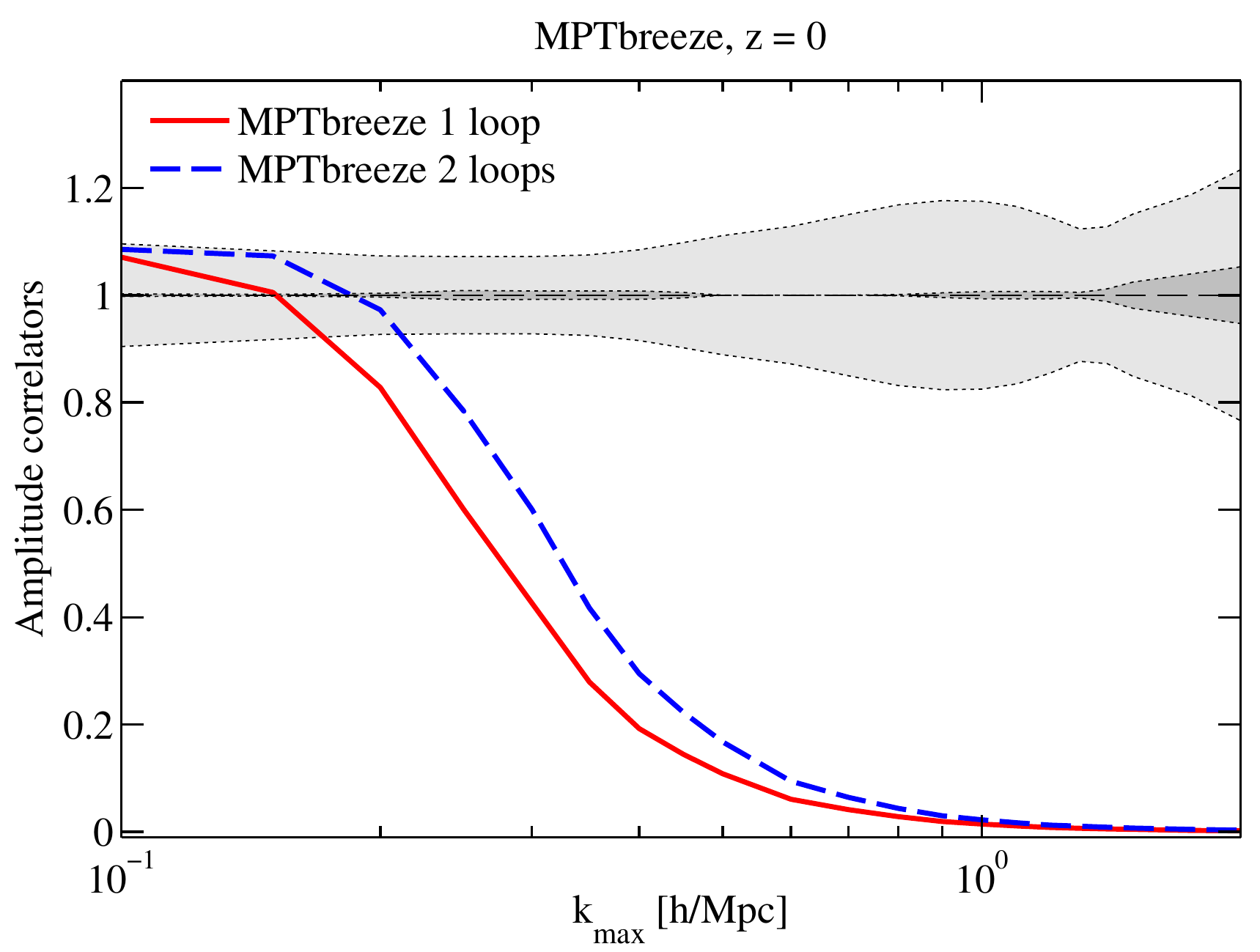} \includegraphics[height=6 cm]{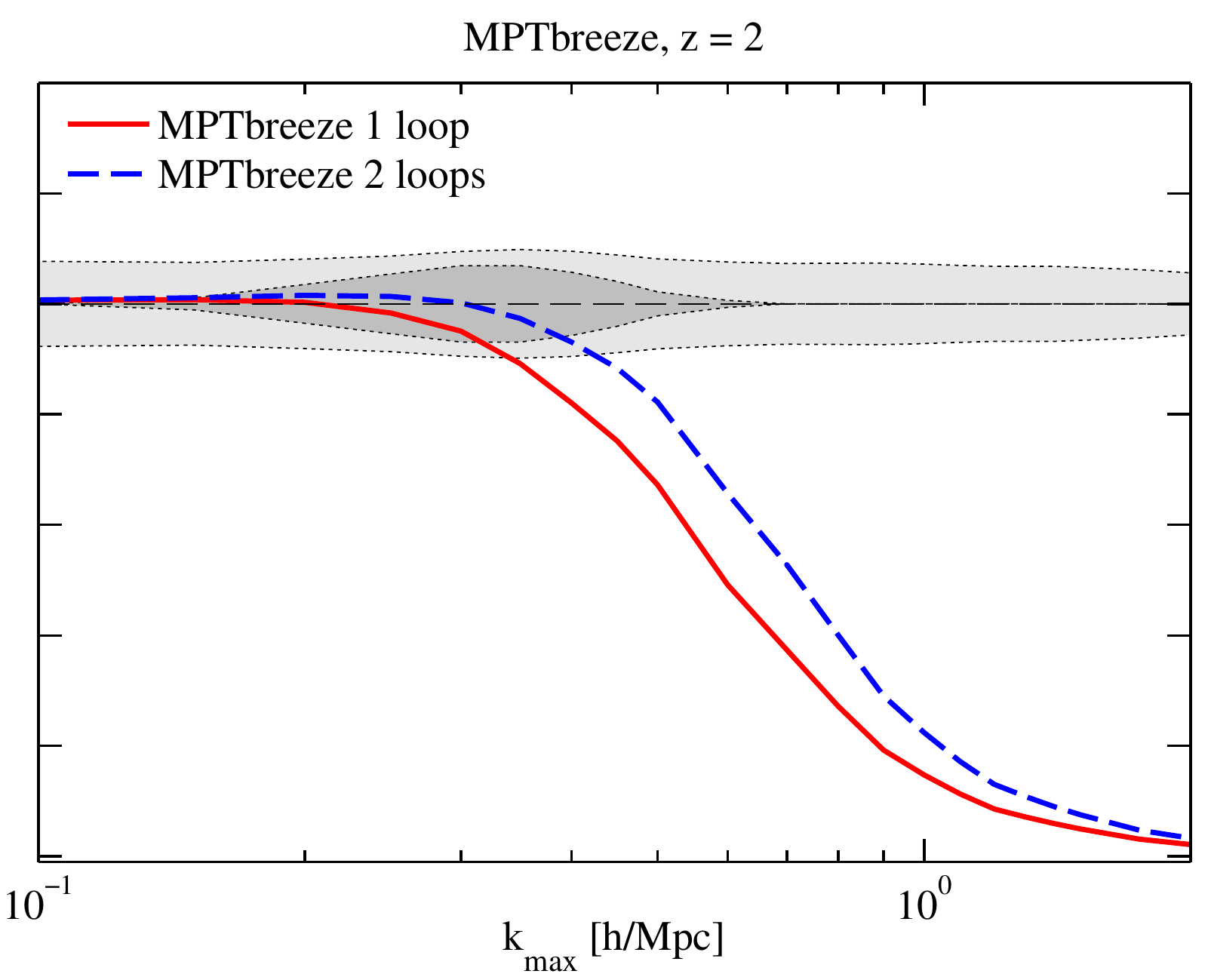} 
\caption{Amplitude correlators between the \textsc{MPTbreeze} model at one and two loops and the three-shape model, at $z=0$ (left) and $z=2$ (right).}
\end{center}
\end{figure}

In the case of the halo model, the features and deficiencies of it can be clearly seen in the same plots. At redshift zero (red line) there is an excess of power on large scales, while this excess is drastically reduced at $z=2$. At this redshift one can see the inaccuracies in the modelling of the two-halo term, which is significantly underestimated, with the halo model giving only around 50\% of the signal in the simulations. The dashed purple line, representing the halo-PT model (combined with EFT) solves the large-scale issue by using EFT on large scales and cutting off the large-scale part of the two-halo term \cite{bislong}.

\subsection{Two-loop results}
We have analysed the two-loop bispectrum in \textsc{MPTbreeze}. Details of the terms that are present in this case and how to take care of divergences are given in the Appendix of Ref. \cite{bislong}. In Fig. 3 we have plotted the amplitude correlators between this model at one and two loops and the three shape model. There is significant gain in going to two loops in this model, as the scale where the theory is decaying is shifting towards nonlinear scales with around 0.1 $h/\mathrm{Mpc}$. This model could be used in the future for an enhanced three-shape model.

\subsection{Primordial non-Gaussianity}
We have investigated the non-Gaussian corrections to the matter bispectrum for the case of local and equilateral types of PNG, by analysing the predictions of the tree-level, one loop SPT, EFT and halo model (Fig. 4). The predictions are very similar to those found in the Gaussian scenario \cite{2017PhRvD..95h3511L} -- the one-loop SPT provides an excessive signal on mildly non-linear scales for both local and equilateral shapes, that can be subtracted using counterterms in EFT. In the case of the halo model the deficit of power is appearing on similar scales to the Gaussian scenario.

\begin{figure}
\begin{center}
\includegraphics[height=6 cm]{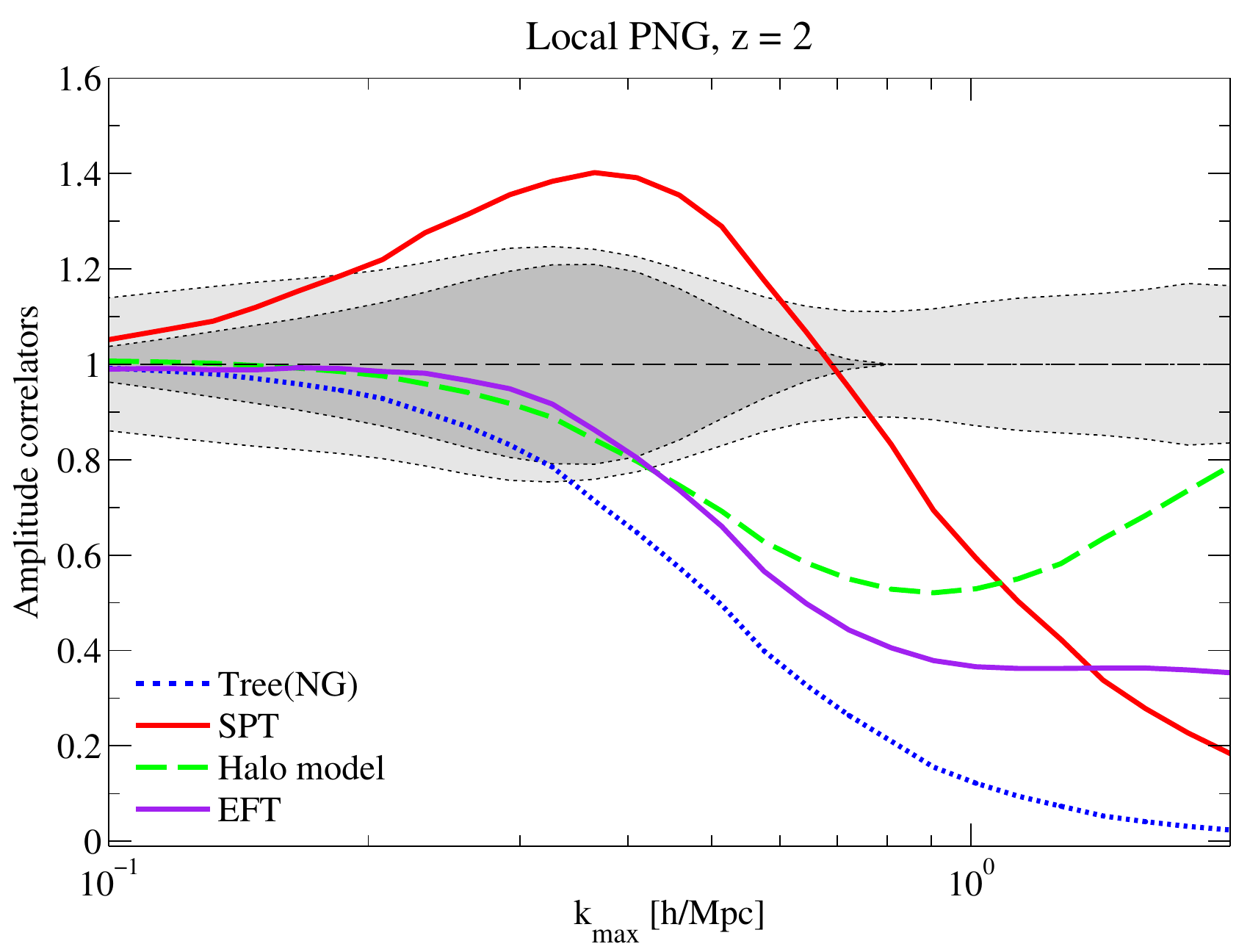} \includegraphics[height=6 cm]{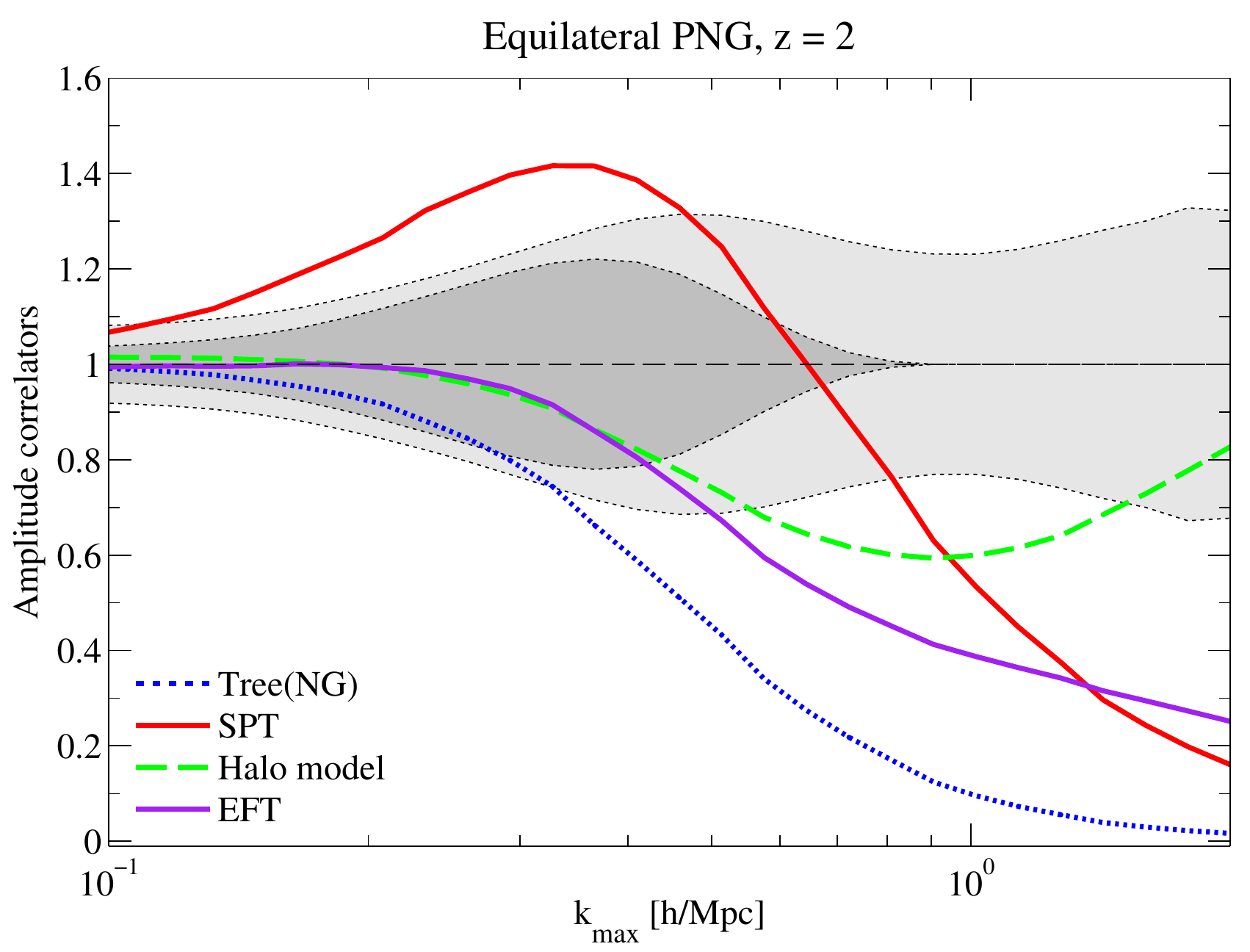} 
\caption{Amplitude correlators between the non-Gaussian corrections to the tree-level bispectrum, one-loop bispectrum, EFT the halo model, and  $\Delta B_{NG}^{3-\mathrm{shape}}$ at $z=2$ for local (left) and equilateral (right) types of non-Gaussianity.}
\end{center}
\end{figure}

\section{Conclusions}

In this paper we have performed a detailed investigation of the matter bispectrum of large scale structure. We have analysed the predictions of perturbation theories and halo models for both Gaussian and non-Gaussian initial conditions. In the Gaussian scenario, we have shown that in the case of perturbation theories the effective field theory of large scale structure (at one loop) extends the furthest into the non-linear regime. This requires nevertheless fitting free parameters to the power spectrum of simulations. The two-loop bispectrum of \textsc{MPTbreeze} provides a competitive option to EFT, being parameter-free.

We have investigated the bispectrum of the halo model and we have shown that each of its three terms 
has a different shape -- a constant shape, squeezed shape and flattened shape -- and we have used this observation to build a new phenomenological, three-shape model that provides an accurate description to the matter bispectrum of LSS by fitting a few parameters. In the process we have identified inaccuracies in the modelling of the two-halo term of the halo model, that account for an overall underestimation of the bispectrum on intermediate scales.

We have also shown that this model can be extended to PNG and that the three shapes are preserved even in that situation.

\section*{Acknowledgements}
First of all I would like to express my gratitude to the organisers of the \textit{Rencontres du Vietnam Cosmology} 2017 Conference, and especially to Professors Jacques Dumarchez and Jean Tr\^{a}n Thanh V\^{a}n,  for inviting me to attend the conference, for their hospitality in Quy Nhon and for creating such a stimulating environment for cosmology.
I would also like to thank Paul Shellard, Tommaso Giannantonio and Marcel Schmittfull for the fruitful collaboration which leaded to Refs. \cite{bislong,2017PhRvD..95h3511L}.

\end{document}